# Light-induced optical orientation of magnetic moments in transition-metal doped hybrid metal halide perovskites

Stanislav Bodnar,[1,+] Jonathan Zerhoch,[1,+] Shangpu Liu,[1] Andrii Shcherbakov,[1] Markus W. Heindl[1], Alexey Sapozhnik,[2] and Felix Deschler[1*].

[1]Physikalisch-Chemisches Institut, Universität Heidelberg, Im Neuenheimer Feld 229, 69120 Heidelberg, Germany

[2]Institute of Physics, LUMES, École Polytechnique Fédérale de Lausanne (EPFL), 1015 Lausanne, Switzerland

\* Corresponding author. Email: felix.deschler@pci.uni-heidelberg.de

[+] Authors contributed equally

Abstract

Using optical orientation to manipulate magnetic moments in matter with light is a key objective in opto-spintronics, however, realizations of such control on ultrafast timescales are limited. Here, we report ultrafast optical control of magnetic moment orientation in magnetically doped metal halide perovskites. Employing intense pulses of circularly polarized light, we inject populations of spin-polarized charge carriers in pristine and manganese-doped $MAPbBr_3$ thin films. Using transient Faraday rotation spectroscopy, we probe the ultrafast magnetic moment dynamics following photoexcitation and find that light-induced magnetization in doped samples is increased by a factor of 10. We attribute this to photoexcited carriers acting on the magnetic moments of manganese dopant-ions via the sp-d exchange interaction, which forces them to align on picosecond timescales. Our findings open new avenues for device structures that use hybrid metal halide perovskites for ultrafast optical manipulation and read-out of magnetic order with the potential for high switching rates.

Introduction

Organic-inorganic metal halide perovskites (HOIPs) have attracted attention due to their extraordinary optical and semiconducting properties, making them promising candidates for device applications. These semiconductors are efficient light absorbers and emitters with high quantum yields [1,2]. Notably, variations in chemical composition, such as halide mixing or implementation of organic groups of different sizes [3-5], enable tuning of the bandgap energy, which is an important design property for light-emitting diodes (LED) [6-8] and photovoltaic applications [9, 10]. Simultaneously, HOIPs can tolerate defects and preserve their optical and electronic properties even at high defect concentrations [11, 12]. Further, HOIPs exhibit strong spin-orbit coupling (SOC), which enables efficient injection of spin-polarized photoexcited charge carriers and can modulate the band structure via the Rashba effect [13, 14]. Due to these properties, HOIPs are also considered as a prospective platform for spintronics applications [15-20]. A promising approach for the application of HOIPs in spintronics is their use in combination with transition metal dopants, such as Mn, Co, or Ni. Neumann et al. reported on the properties of diluted Mn-doped 2D perovskites [21]. It was found that the presence of Mn-ions in the 2D perovskite lattice affects the excitonic properties of the host material, such as enhanced circularly polarized photoluminescence (PL) in the presence of external magnetic fields. Despite these promising properties of magnetically doped perovskites [22, 23], the optical orientation and readout of magnetic order in these materials have not been realized, yet.

Magnetic and optical properties of Mn-doped $MAPbBr_3$ hybrid perovskites

To study the all-optical control of the magnetic moments of dopants in HOIPs we chose the well-studied 3D hybrid perovskite $MAPbBr_3$ (MA = $CH_3NH_3$, methylammonium) (see inset Fig.1a). We doped $MAPbBr_3$ thin films with high concentrations of $Mn^{2+}$ dopants (see details on preparation in the Supplementary Information (SI)). X-ray diffraction (XRD) θ/2θ-scans show the same crystal phase for pristine $MAPbBr_3$ and for 50 % Mn-doped thin films with a slightly modified lattice constant upon Mn-doping (Fig. S1). The resulting concentration of $Mn^{2+}$ dopants in the film was estimated by energy-dispersive X-ray spectroscopy (EDX), and is around 50 % with respect to the concentration of Pb atoms

(Supplementary Fig. S2 and Table T1). To investigate the magnetic properties of Mn-doped MAPbBr$_3$ thin films, we performed experiments using a superconducting quantum interference device (SQUID), MPMS3 Quantum design (See. Fig. 1a,b). Our SQUID data show the paramagnetic behavior of Mn-doped MAPbBr$_3$ films in the temperature range 2 - 293 K without any sign of magnetic phase transitions, which underlines the absence of magnetic clusters such as MnO, MnBr$_2$, or metallic Mn. Further, we examined our samples using PL and absorption spectroscopy (Fig. 1c). The optical characterization revealed shifts in the maxima of the PL-spectra and of the absorption edges by ~60 meV towards the low energy side for the Mn-doped samples.

The thickness and morphology of the samples were examined using atomic force microscopy (AFM) and optical microscopy (Fig. S3). The pristine sample forms a continuous polycrystalline thin film with an average thickness of about 200 nm. The Mn-doped sample shows an island-like morphology pattern with an average area coverage of about 20 %. The average thickness of the islands is about 180 nm which is comparable to the pristine MAPbBr$_3$ sample (full details can be found in the SI). In conclusion, the material characterization shows the loading of perovskite films with high concentrations of Mn$^{2+}$ dopants. The rather small changes in diffraction patterns suggest that the dopant atoms are present as interstitials in the MAPbBr$_3$ crystal lattice [24], or as nanoscale clusters below the resolution limit of EDX. We note that, as relevant for our study, in either case, strong interactions between photoinduced charge carriers and dopant moments will arise.

Photoinduced magnetic moment alignment in Mn-doped MAPbBr$_3$

To investigate the spin dynamics of the photoexcited charge carriers of Mn-doped and pristine MAPbBr$_3$, we used transient Faraday rotation (FR) pump-probe spectroscopy. Laser pump pulses (photon energy 2.41 eV, pulse duration 300 fs) were circularly polarized, while probe pulses at 1.77 eV, with energy below the bandgap of MAPbBr$_3$, were linearly polarized.

The absorption of circularly polarized light in MAPbBr$_3$ leads to the injection of spin-polarized charge carriers (SPCCs) from the valence into the conduction band [19]. A schematic picture of the optical transitions in MAPbBr$_3$ can be found in the SI (Fig. S8).

The injection of SPCCs can be considered as the photoexcitation of a transient effective magnetization $M_{eff}$ which is created by the ensemble of the SPCCs magnetic moments. This transient effective magnetization induces rotation of the polarization plane of the linearly polarized probe pulse due to the Faraday effect [25]. The value of the transient FR angle ($\theta_F$) is linearly proportional to the amplitude of $M_{eff}$ induced by the pump. We record $\theta_F$ as a function of the delay time in pristine and 50 % Mn-doped MAPbBr$_3$ samples measured for pump fluences of 20 µJ/cm$^2$ and 100 µJ/cm$^2$ (Fig. 2a). The higher excitation fluences lead to a higher $\theta_F$ at zero-time delay ($FR_{t=0ps}$), where the maximum of the transient signal occurs. This agrees with pump pulses of a higher fluence injecting a higher number of SPCCs, and hence causing a higher $M_{eff}$. Unexpectedly, for a pump fluence of 20 µJ/cm$^2$, the amplitude of the $FR_{t=0ps}$ is higher for pristine samples, whereas for 100 µJ/cm$^2$, the amplitude is higher for the Mn-doped sample. We attribute the additional transient Faraday signal to the light-induced alignment of the manganese magnetic moments.

To understand how the light-induced transient magnetic moments of manganese ions and of SPCCs behave as a function of pump fluence we measured $FR_{t=0ps}$, which corresponds to the maximum total effective magnetization, for a wide range of fluences for the Mn-doped and pristine samples. Fig. 2b shows $FR_{t=0ps}$ as a function of pump fluence for 1 %, 25 %, and 50 % Mn-doped and pristine MAPbBr$_3$. In addition to the SPCCs, the Mn-doped sample in the higher fluence regime gains an additional light-induced magnetization which originates from the manganese spins and the contribution of the manganese spins starts to play a dominating role.

To quantify the effective photoinduced transient magnetization which is associated with aligned manganese dopant moments, $M(\text{Mn})$, we used the following equation:

$$M(\text{Mn}) = \frac{FR(\text{Mn},\ t=0\ \text{ps})}{GSB(\text{Mn},\ t=0\ \text{ps})} / \frac{FR(\text{Prisitne},\ t=0\ \text{ps})}{GSB(\text{Prisitne},\ t=0\ \text{ps})} \tag{1}$$

where $FR(\text{Mn/Pristine},\ t=0\ \text{ps})$, is the $\theta_F$ at the delay time $t = 0$ ps for Mn-doped and undoped MAPbBr$_3$, $GSB(\text{Mn},\ t=0\ \text{ps})$ and $GSB(\text{Pristine},\ t=0\ \text{ps})$ are the $\Delta T/T$ transient absorption (TA) signals at the energies which correspond to the ground state bleaches (Fig. S9) for Mn-doped and pristine samples, respectively. Here, the $\Delta T/T$ at the spectral position of the *GSB* allows us to consider differences in how efficiently charge carriers can be generated by circularly polarized light for Mn-doped and for pristine samples. We stress that the *M*(Mn) reflects the relative transient light-induced magnetization of manganese dopants and is not the same quantity that was measured using SQUID.

Using equation 1, we extracted the part of the effective photoinduced transient magnetization which is associated with aligned manganese spins *M*(Mn) as the function of the pump fluence for different sample temperatures (Fig. 2c) for the 50 % Mn-doped sample. We note that the pump fluence can be read as the strength of the $M_{eff}$ induced by SPCCs. We find that the fluence dependence of optically-oriented manganese spins follows a Brillouin function with J = 5/2. This indicates that the optically-induced SPCCs act on the Mn-dopants in the same way as external magnetic fields, yet, on ultrafast timescales.

Further, we measured the *M*(Mn) as a function of pump fluence for different temperatures (Fig. 2d). For lower temperatures, at the same fluences, we observe a higher effective dopant magnetization *M*(Mn). The Mn-doped sample is a paramagnetic material, which means that manganese atoms do not form an ordered magnetic state and the magnetization of manganese ions induced via the interaction with SPCCs should also follow the paramagnetic Curie law, similar to the SQUID data [26]. Similarly, to the temperature dependence of the magnetization for Mn-doped samples obtained via SQUID, we observe a drop in *M*(Mn) with temperature rise (Fig. 2d) with a trend for an increase in the light-induced manganese magnetization *M*(Mn) for higher fluences. This dependence also revealed the paramagnetic-like behavior of the light-induced manganese magnetization. We extracted *M*(Mn) for 1 %, 25 %, and 50 % Mn-doped samples for high pump fluences where the signal saturates (Fig. 2e). We find that the contribution of light-polarized manganese spins to the transient Faraday signal scales linearly with the concentration of Mn-doping, in agreement with ultrafast optically aligned paramagnetic manganese spins.

To test the impact of non-magnetic doping on the transient Faraday signal due to possible strain impact on the band structure [27] we also have performed transient Faraday rotation measurements on non-magnetically doped MAPbBr$_3$, particularly Mg-doped. The XRD data confirms the same crystal phase of Mg-doped samples as in the case of the pristine samples. The Mg-doping leads to the presence of intrinsic strain which can be seen from XRD θ/2θ scans (Fig. S1). We compared the signal obtained in pristine, Mn-doped, and Mg-doped samples for the highest fluence of pump pulses of 900 µJ/cm$^2$ due to the strongest additional transient Faraday signal. After normalization on the absorption at the pump wavelength (2.41 eV) and by the material volume covered by the probe pulses (more details SI) and found that the signal $FR_{t=0ps}$ is similar for the Mg-doped and the pristine samples (Fig. 2f) while the Mn-doped sample shows an approximately 10 times stronger signal. Thus, non-magnetic doping can be ruled out as the source of the enhanced transient Faraday signal.

Photoinduced Spin Dynamics in Mn-doped MAPbBr$_3$

To gain insights into the underlying mechanism of SPCCs interactions with dopant moments, we analyzed the kinetics of the normalized transient Faraday angle to study how the optical alignment of paramagnetic dopants affects spin-relaxation dynamics (Fig. S13) and found that doped samples show longer spin-lifetimes. To understand the origin of this behavior, we plot the normalized kinetics of the effective light-induced magnetization of manganese atoms $M$(Mn) for different pump fluences (Fig. 3a). We find that the normalized $M$(Mn) signal shows a rise time $\tau_R$ which we quantify by the delay times at which the maximum in $M$(Mn) occurs. We attribute the rise time of the additional transient signal to the alignment of manganese spins induced by SPCCs. The rise times further depend inversely on the pump fluence (Fig. 3b). This agrees with a scenario where the probability for manganese ions to interact with SPCCs via the sp-d exchange interaction grows linearly with the number of SPCCs. The rise time $\tau_R$ can be considered as the time required for manganese ions to align along the spin polarization direction via the sp-d exchange interaction with SPCCs. This time is defined by the probability of the SPCCs interacting with manganese ions via the sp-d exchange and therefore should depend on the number of SPCCs and manganese atoms. The schematic picture of the alignment process of manganese spins for the case of small and high fluences is shown in Fig. 3b (inset).

Additionally, we extracted the rise time $\tau_R$ as the function of the manganese doping level. The rise time $\tau_R$ of the manganese spins alignment via the sp-d exchange interaction should further be inversely proportional to the number of manganese atoms at fixed pump fluences. This can be seen in Fig. 3c which shows the rise time as the function of the manganese content for 1 %, 25 %, 50 %, and 75 % of manganese doping at the fixed pump fluence of 900 µJ/cm$^2$ measured at room temperature. The rise time $\tau_R$ in this case is inversely proportional to the number of manganese atoms as one should expect in the case of sp-d exchange interaction.

We also examined the spin dynamics measured with transient Faraday rotation with respect to the dynamics of photoexcited charge carriers excited with linearly polarized pump and probe pulses. Figure S13 shows the transient Faraday signal normalized by the *ΔT/T*-value taken at the energies corresponding to the *GSB*. This figure shows how fast the total transient magnetization decays compared to the rate at which electrons relax from the excited state. For the first 2 ps the signals are quite similar while for later delay times, the Mn-doped sample shows longer spin dynamics. This phenomenon can be seen as the presence of some magnetization originating not from the spin-polarized electrons but likely from aligned manganese spins. Figure S11 shows the spin lifetimes as a function of fluence and temperature. The Mn-doped sample shows a longer spin lifetime for the whole range of fluences and temperatures.

Discussion

We can rationalize the strong FR signal in Mn-doped samples by considering that Mn$^{2+}$ ions have 5 unpaired electron spins which can be aligned by the photoinduced populations of SPCCs following a circularly polarized pump [28-90], while in pristine samples SPCCs are the only source of transient magnetic moments. This phenomenon was shown in Mn-doped III-V semiconductors, in which SPCCs can be used to align manganese spins via the sp-d exchange interaction [28-30]. Thus, the total transient Faraday signal in the Mn-doped MAPbBr$_3$ consists of two parts: the transient signal associated with light-induced SPCCs, and manganese spins aligned by SPCCs. We assign the sp-d exchange interaction between manganese spins and optically-oriented SPCCs as the responsible mechanism for the optical manganese spin alignment.

Notably, the observed alignment of manganese spins within a few picoseconds due to the sp-d exchange mechanism is a very fast process considering typical characteristic timescales for the light-induced manipulation of magnetic order in ferromagnets or antiferromagnets. In both cases, the speed with which spins can be manipulated is limited by the spin precession modes. In ferromagnets, it has been claimed that the light-induced magnetization switching can be as fast as a few tens of picoseconds [31]. In antiferromagnets, this limit can be lowered to 1 ps due to antiferromagnetic THz spin dynamics [32]. In the case of Mn-doped MAPbBr$_3$, we show that control of manganese spins can be achieved as fast as 2 ps for fluences above 200 µJ/cm$^2$. Such fast alignment of the manganese magnetic moments is enabled by the exchange nature of the interaction between SPCCs and d-electrons of Mn, which has been reported to be a fast process with charateristic times varying between few dozens of femtoseconds [33] to few ps [28-30].

The interaction of SPCCs with magnetic impurities has been studied for a range of dilute magnetic semiconductors under continuous wave optical excitation. For instance, Akimov et al. [28] investigated the effect of optical orientation of manganese spins in diluted Mn-doped GaAs and have shown that injection of photoexcited SPCCs can align manganese spins. We believe that a similar mechanism of sp-d exchange alignment of manganese spins occurs in our Mn-doped MAPbBr$_3$. Yet, we find that such effects can occur on the ultrafast timescales in a previously inaccessible regime of few picoseconds.

<u>Conclusion.</u>

In conclusion, we report ultrafast all-optical control of magnetic dopants' spins in metal halide perovskites using circularly polarized light. We find that the photoexcited population of SPCCs interacts with manganese spins via the sp-d exchange interaction. This interaction leads to the alignment of manganese spins along the direction of the magnetic moment of the photoexcited SPCCs. Our findings show, that HOIPs hold unexpected potential as solution-processable systems for efficient all-optical spin control and spin injection. We expect that this property can also be exploited in heterostructures with established high-performance ferromagnets [34, 35] and emerging antiferromagnetic [36-38] materials for electrical read-out of the optically-set spin state. The ultrafast nature of the observed mechanism holds the potential for achieving unprecedented rates of magnetization switching for computing and information processing.

**Acknowledgements:** The authors acknowledge Prof. Dr. Rüdiger Klingeler for helpful scientific discussion and for access to the MPMS system. The authors acknowledge Dr. Penelope Monkhouse for proof-reading wording and grammar of the manuscript.

**Funding:** This project received funding from the European Research Council (ERC Starting Grant agreement no. 852084 — TWIST).

**Authors information:** These authors contributed equally: Stanislav Bodnar, Jonathan Zerhoch.

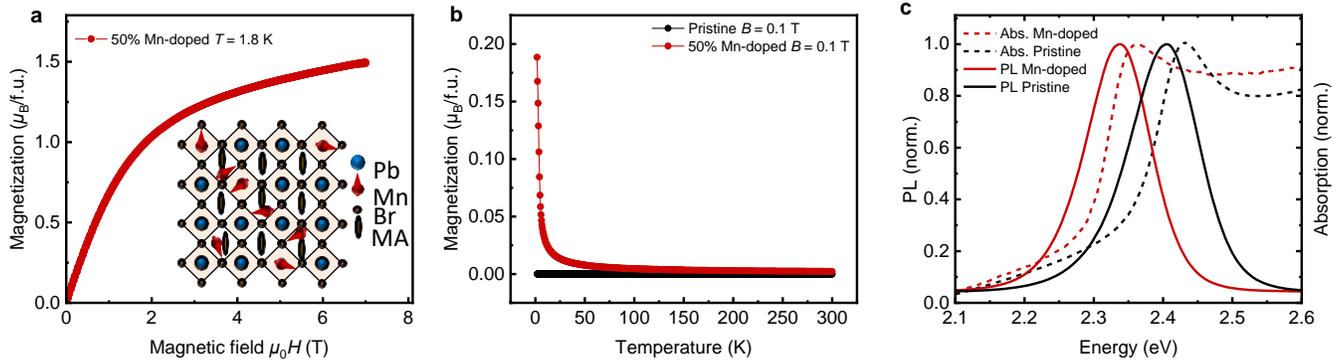

*Fig. 1. Optical and magnetic properties of (Mn-doped) MAPbBr₃.* **(a)** Magnetization measurement of 50 % Mn-doped MAPbBr₃ at 1.8 K. Inset shows a schematic picture of the Mn-doped MAPbBr₃ materials. **(b)** Magnetization of the 50 % Mn-doped (red) and pristine (black) MAPbBr₃ thin films as a function of temperature measured at an external magnetic field of $B = 0.1$ T. **(c)** PL and absorption spectra measured at room temperature for 50 % Mn-doped and pristine MAPbBr₃ thin films. Solid lines correspond to the PL spectra and dashed lines correspond to the absorption spectra.

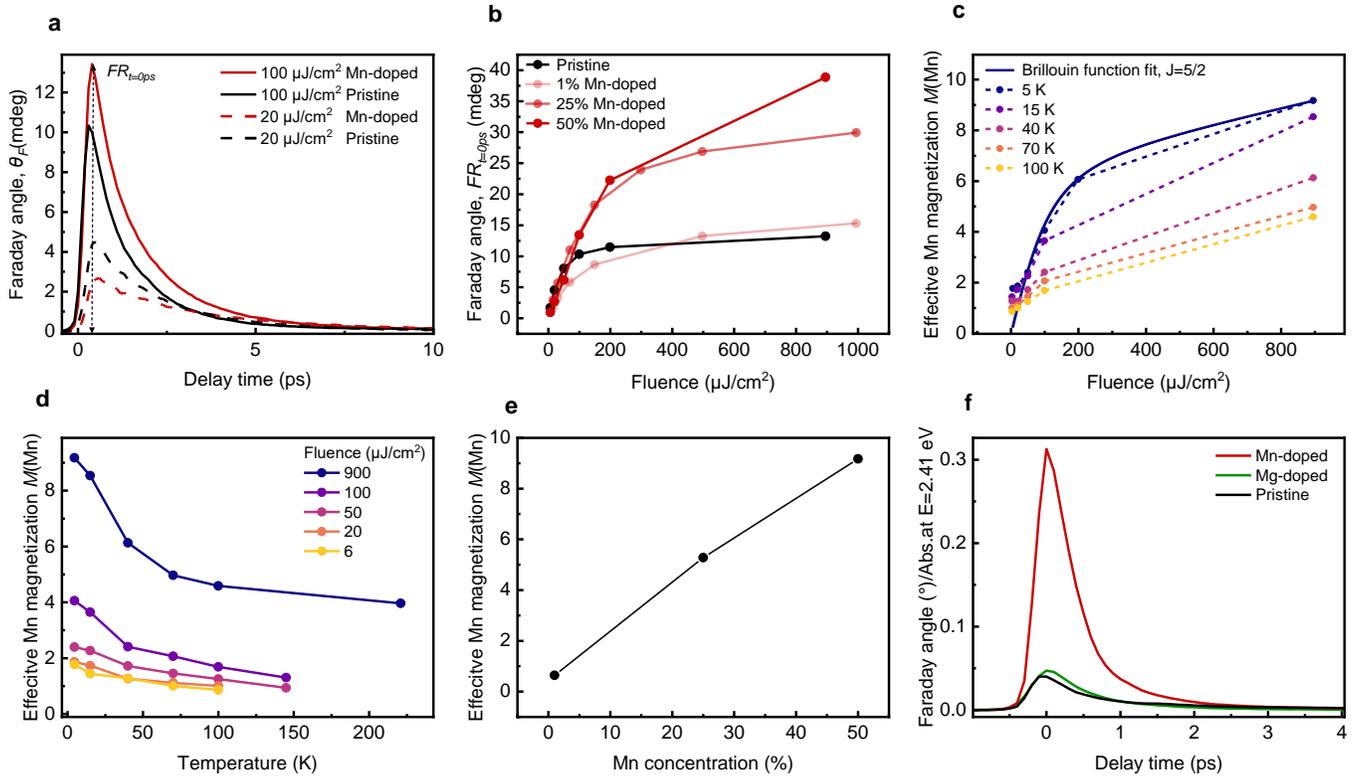

***Fig. 2. Ultrafast all-optical control of Mn-dopants magnetic moments**. **(a)** Transient Faraday angle measured at 5 K for 50 % Mn-doped and pristine MAPbBr$_3$ for two fluences of the excitation 20 μJ/cm$^2$ and 100 μJ/cm$^2$. **(b)** Faraday angle at delay time t = 0 ps measured for 25 %, 50 %, 1 % Mn-doped and pristine MAPbBr$_3$ thin films as a function of pump fluence. **(c)** Light-induced effective magnetization M(Mn) of Mn-dopants in the 50 % Mn-doped thin film measured as the function of pump fluence for different temperatures. M(Mn) was calculated using equation (1) from the main text. Dots indicate experimental data points; solid line indicates the fit of the experimental data by a Brillouin function with J = 5/2 for 5 K. **(d)** Light-induced magnetization M(Mn) as the function of temperature measured for different pump fluences. **(e)** Effective light-induced magnetization M(Mn) as a function of Mn concentration for a pump fluence of 900 μJ/cm$^2$. **(f)** Transient Faraday rotation angle as a function of the delay time normalized by absorption at the photon energy of the pump pulses (E=2.41 eV) and the excited material volume.*

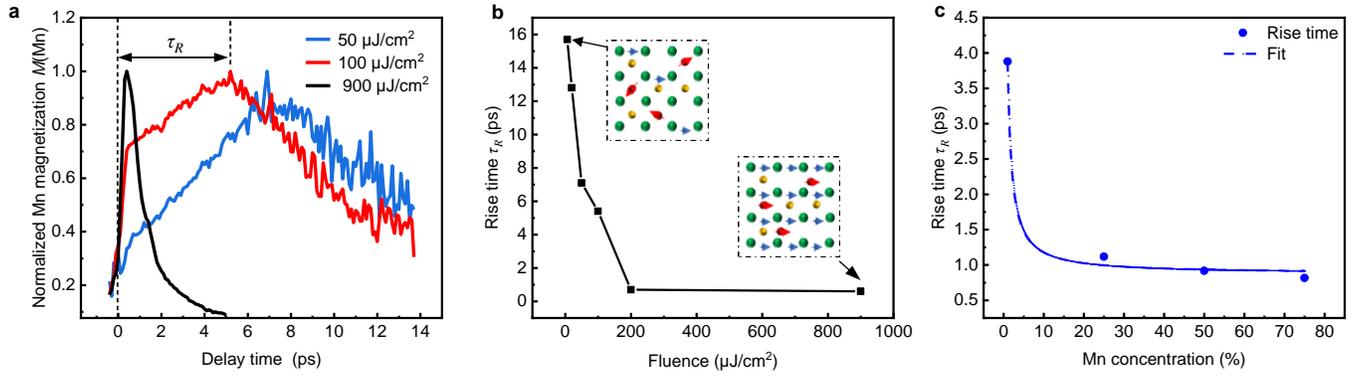

*Fig. 3.* **Spin dynamics in Mn-doped MAPbBr$_3$.** *(a)* Normalized effective magnetization of Mn $M(Mn)$ as a function of delay time measured for different pump fluences. Arrows indicate extracted rise times $\tau_R$ for different fluences. *(b)* Rise time as a function of pump fluence measured for 50 % Mn-doped MAPbBr$_3$ at 5 K. *(c)* Rise time as a function of the Mn doping concentration measured for 900 µJ/cm$^2$.

# Light-induced optical orientation of magnetic moments in transition-metal doped hybrid metal halide perovskites

Stanislav Bodnar,[1,+] Jonathan Zerhoch,[1,+] Shangpu Liu,[1] Andrii Shcherbakov,[1] Markus W. Heindl[1], Alexey Sapozhnik,[2] and Felix Deschler[1*].

[1]Physikalisch-Chemisches Institut, Universität Heidelberg, Im Neuenheimer Feld 229, 69120 Heidelberg, Germany

[2]Institute of Physics, LUMES, École Polytechnique Fédérale de Lausanne (EPFL), 1015 Lausanne, Switzerland

* Corresponding author. Email: felix.deschler@pci.uni-heidelberg.de

[+] Authors contributed equally

## 1. Sample preparation

Pb(OAc)$_2$·3H$_2$O, MABr, DMF, and Mn(OAc)$_2$·4H$_2$O, were purchased from Sigma-Aldrich, Mg(OAc)$_2$·4H$_2$O from Thermo Scientific. All chemicals were used without further purification. For a general fabrication of MAPbBr$_3$ thin films, 0.068 g MABr and 0.076 g Pb(OAc)$_2$·3H$_2$O were dissolved in 1 mL DMF at 80 °C for 30 min. Then, the precursor solution was spin-coated on glass substrates with a rotation speed of 4000 rpm for 35 s, after which they were annealed at 80 °C for 30 min. The whole process was performed in a nitrogen-filled glovebox. For Mn$^{2+}$/ Mg$^{2+}$ doped MAPbBr$_3$ thin films, all the procedures were similar, except for replacing the same molar amount of Pb(OAc)$_2$·3H$_2$O with Mn(OAc)$_2$·4H$_2$O / Mg(OAc)$_2$·4H$_2$O. The obtained perovskite thin films were stored inside the glovebox for further characterization.

## 2. XRD characterization and energy-dispersive X-ray spectroscopy (EDX).

The X-ray diffraction (XRD) scans for Mn-doped, Mg-doped, and the pristine thin films were obtained using an X-ray diffractometer (Rigaku SmartLab) with Cu K$_\alpha$ radiation with a wavelength of 1.5406 Å. Fig. S1. shows the XRD scans for the 50% Mn-doped and pristine thin films.

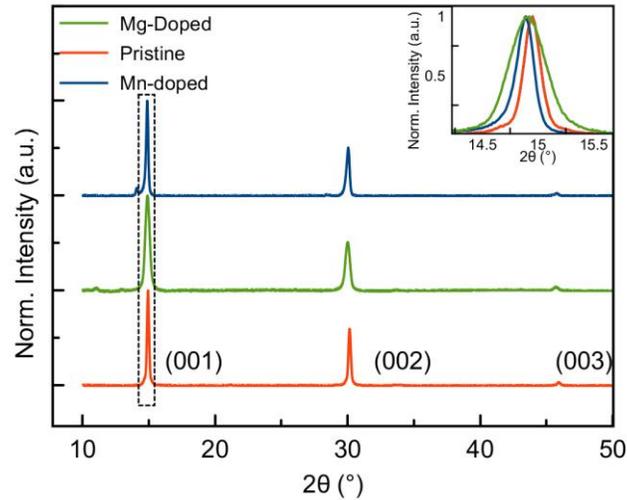

**Fig. S1. XRD data**. *XRD diffraction scans of the pristine (red line), 50% Mg-doped MAPbBr$_3$ (green line), and 50% Mn-doped MAPbBr$_3$ thin films (blue line). The insert shows the zoomed-in region around the (001) peak.*

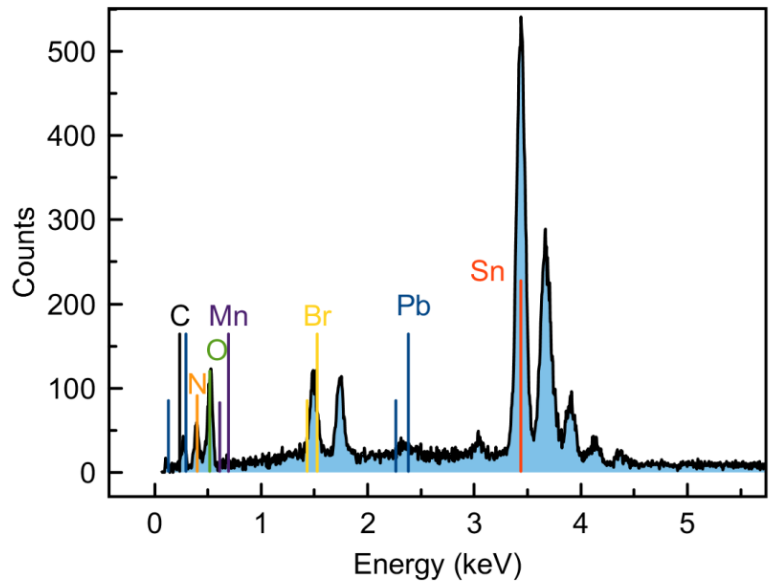

**Fig. S2. EDX spectroscopy**. *EDX spectrum of 50% Mn-doped MAPbBr$_3$ thin film.*

To investigate the atomic composition of Mn-doped MAPbBr3 thin films experiments on EDX spectroscopy were performed. Measurements on EDX were performed with the help of a Bruker XFlash6130 energy-dispersive X-ray spectrometer, mounted in a Zeiss EVO MA10. For this experiment, a 50 % Mn-doped MAPbBr3 thin film was spin-coated onto an FTO substrate for better conductivity to prevent charging of the sample. The results on EDX can be found in the SI (see Fig. S2 and Table T1).

**Supporting Table T1:** Atomic ratios and mass ratios obtained by the series fit deconvolution.

| Element | Atomic No. | Netto | Mass [%] | Atom [%] | Abs. error [%] (1 sigma) | Rel. error [%] (1 sigma) |
|---|---|---|---|---|---|---|
| Carbon | 6 | 80 | 1,64 | 5,71 | 0,11 | 6,52 |
| Nitrogen | 7 | 194 | 4,15 | 12,42 | 0,22 | 5,38 |
| Oxygen | 8 | 1049 | 20,09 | 52,62 | 1,02 | 5,05 |
| Fluorine | 9 | 51 | 1,44 | 3,17 | 0,10 | 6,91 |
| Manganese | 25 | 78 | 0,82 | 0,62 | 0,05 | 5,68 |
| Bromine | 35 | 54 | 5,50 | 2,88 | 0,16 | 2,96 |
| Tin | 50 | 19079 | 60,64 | 21,41 | 1,63 | 2,69 |
| Lead | 82 | 71 | 5,74 | 1,16 | 0,17 | 2,94 |
| | | **Sum** | | 100,00 | 100,00 | |

### 3. Microscopy characterization of the Mn-doped, Mg-doped and pristine MAPbBr$_3$ thin films.

The thickness and morphology of the thin film samples were examined using atomic force microscopy (AFM) and optical microscopy measurements. Microscopy images and AFM scans are shown in the Fig. S3. To obtain the topographical view of the investigated samples, AFM images were obtained at room temperature using commercial AFM setup (Neasnom, Neaspec, Attocube GmbH). The measurements were performed in the tapping mode with a tapping amplitude of 55 nm and a setpoint value of 80 %. The

PtIr-coated Si tip (NCPt, Nanoworld) had a resonance frequency of around 275 kHz. The pristine thin film forms a continuous polycrystalline thin film with an average thickness of about 200 nm. The Mn-doped thin film shows an island-like morphology pattern with an average area coverage of about 20 %. The average thickness of the islands is about 180 nm which is comparable to the pristine MAPbBr$_3$ thin film.

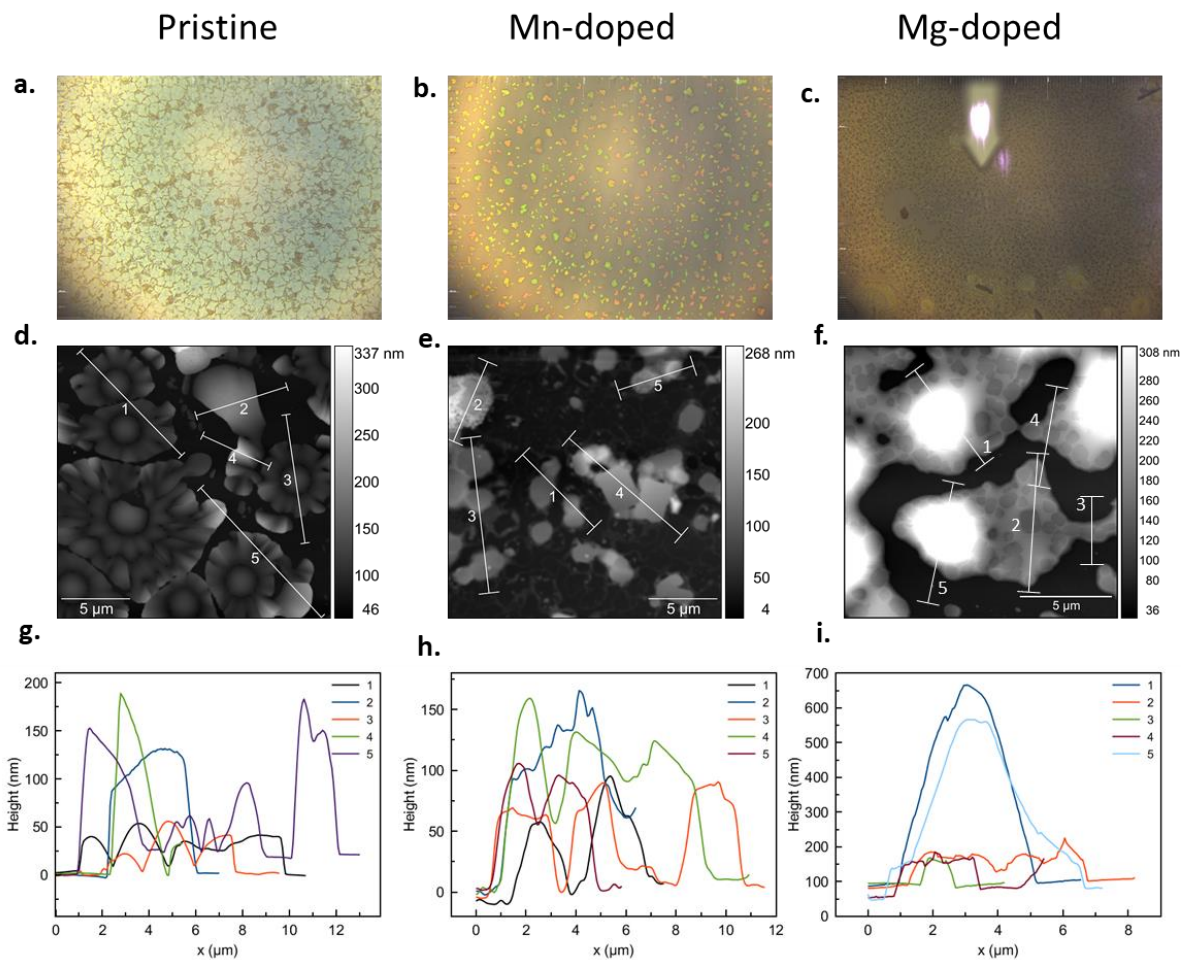

*Fig. S3. Morphology structure investigation. a, b, c) Microscopy images of the pristine, 50% Mn-doped and 50% Mg-doped MAPbBr$_3$ thin films respectively. d, e, f) Atomic force microscopy map of the pristine, 50% Mn-doped, and 50% Mg-doped MAPbBr$_3$ thin films respectively. g, h, i) Height profiles of the samples obtained along the lines under the numbers at the figures d, e and d respectively.*

The evaluation of the occupied and unoccupied areas was done by doing binary operations using ImageJ software on the microscopy images. Fig. S4. shows examples of the binary and direct images of the Mn-doped sample. This data in combination with the AFM data allows us to estimate the average volume of the material per probe beam spot. To estimate the average volume of the material we used the mean height of the islands multiplied by the material-occupied area and normalized by the effective volume of the pristine $MAPbBr_3$ in such way that, the effective volume of the pristine $MAPbBr_3$ is equal to 1.

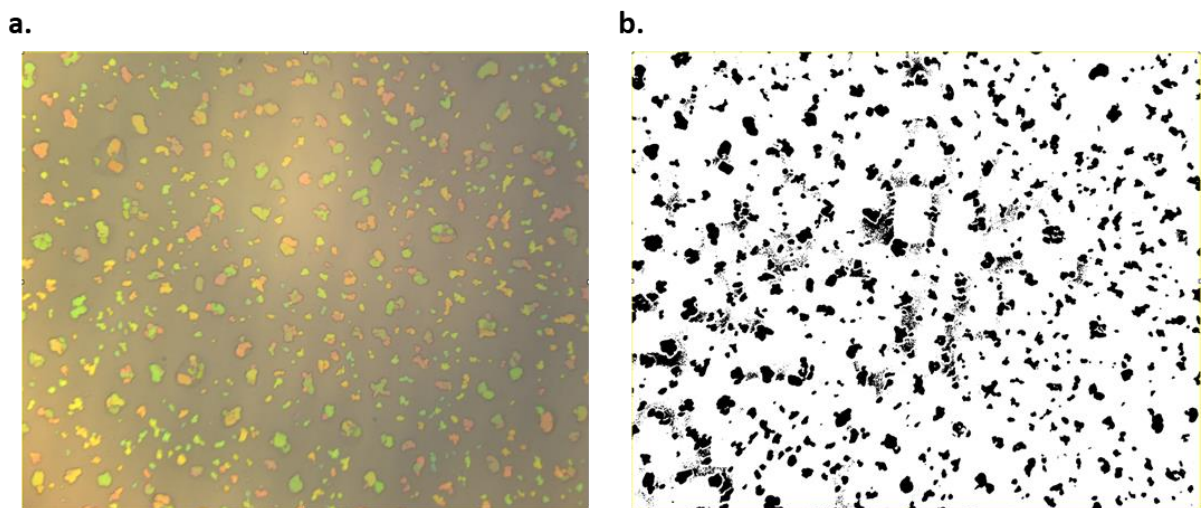

**Fig. S4. Microscopy binary images.** *a) Optical microscopy image of the Mn-doped $MAPbBr_3$ thin film. b) Binary image of the same region.*

Spatially resolved confocal photoluminescence measurements were performed with a customized commercial confocal microscope (Neaspec, Attocube GmbH) with the tunable Continuous wave laser (C-Wave, Hübner Photonics GmbH) for PL excitation and CMOS camera-based spectrometer (Zyla + Kymera, Andor Oxford Instr.). The 460 nm laser beam was spatially filtered, expanded to a diameter of 2 cm, and focused with the off-axis parabolic mirror onto the sample plane. The PL from this excited spot was then collected with the long working distance Objective (N.A.= 0.42) and focused again into the single mode fiber of which the entrance was working as the pinhole. After the fiber, the PL was cleaned from the excitation laser with a 495 nm long-pass filter and finally detected either with the camera for spectrally resolved measurements or photodiode for intensity-resolved spatial maps. Fig. S5. shows the PL intensity map of the Mn-doped sample which has an island-like morphology. The color map reflects the total intensity of the PL. It can be seen that the PL is only visible in the island-like regions while in the part of the sample where there is no material the signal is zero. Additionally, we measured the spectral characteristics of the PL signal acquired from the different regions on the sample. Different islands show the same spectra with the same linewidth and peak position.

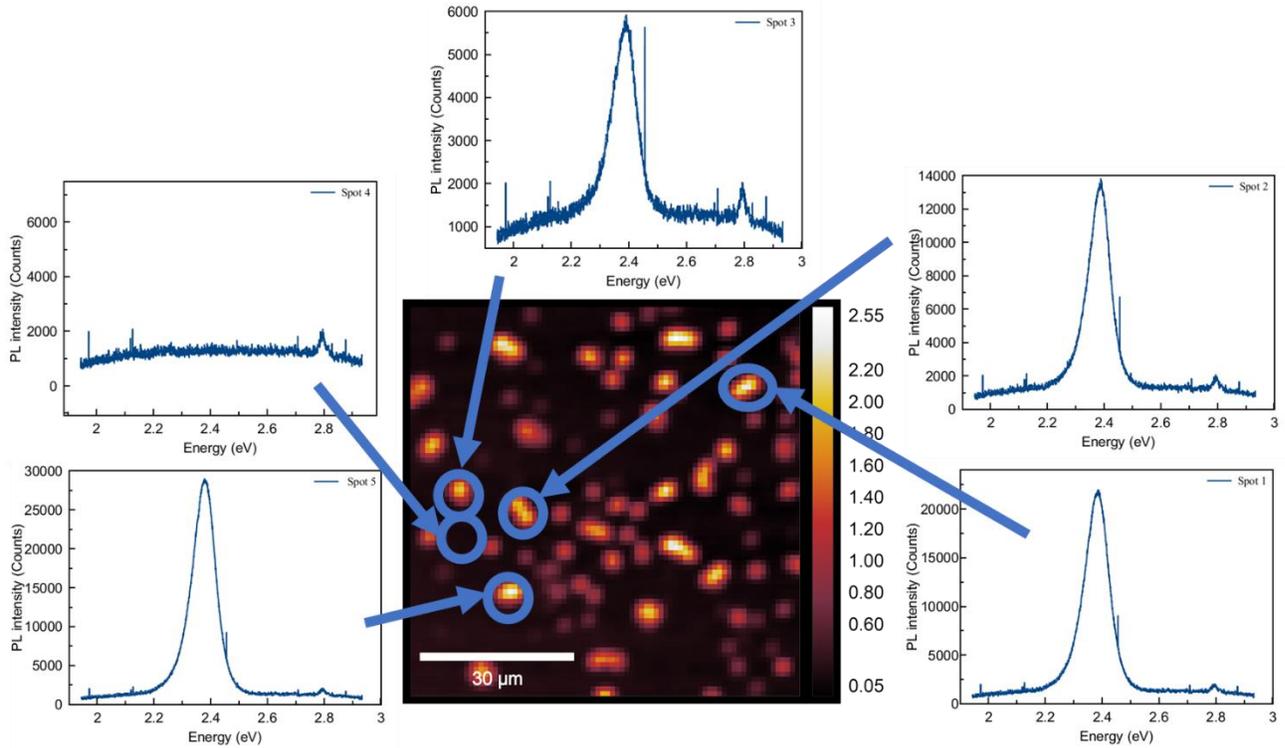

**Fig. S5. Photoluminescence microscopy.** *Confocal PL mapping obtained from the Mn-doped thin film with the spectra at different spots on the sample. The color bar shows relative PL intensity.*

## 4. UV-Vis Absorption spectroscopy characterization.

Fig. S6 shows the absorption spectra of the Mn, Mg, and pristine MAPbBr$_3$ thin films measured in the UV-visible spectral range. To measure absorption spectra, we used a Shimadzu UV-1800 UV-Vis spectrometer which can cover a spectral range between 190 and 1100 nm.

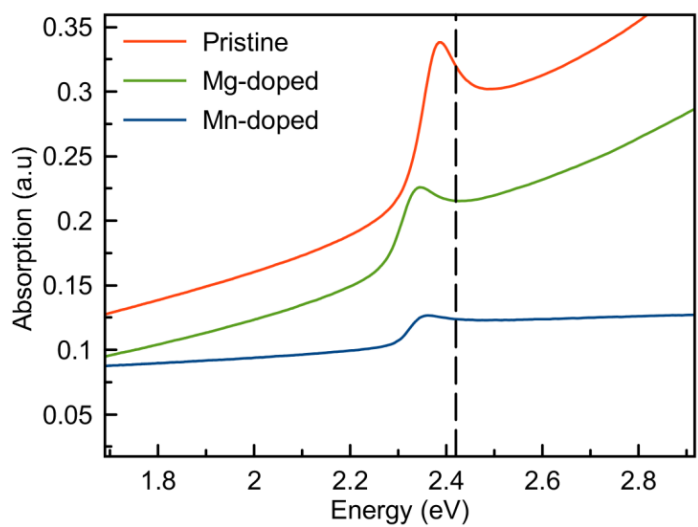

***Fig. S6. Absorption spectra.*** *Absorption spectra measured for Mg-doped, Mn-doped and pristine MAPbBr$_3$ thin films. The dashed line indicates the energy position of the pump used for transient Faraday measurements (E=2.41 eV).*

## 5. Transient Faraday signal in Mg-doped MAPbBr$_3$.

The experiments on transient pump-probe Faraday spectroscopy were conducted at the optical setup where samples were located inside the closed cycle Cryostation (Montana Instruments) which allows for temperature settings in the range between 5 K – 350 K. Transient Faraday pump-probe experiments were performed at a 16.666 kHz laser repetition rate (Pharos by Light Conversion). The pump beam with 2.41 eV energy was guided through the delay stage (Newport DLS325) and a chopper wheel to change the repetition rate to 8.333 kHz. Before hitting the sample, the pump beam was circularly polarized by an achromatic Thorlabs quarter waveplate specified for the spectral region between 300 – 800 nm. The linearly polarized probe pulse was set to 1.77 eV energy and was spatially overlapped with the pump pulse on the sample. After the sample, the probe beam was guided through a Wollaston prism which splits the probe into two components with linear vertical and horizontal polarization. These beams were focused on the balanced photodetector (PDB210A) by Thorlabs. Differential signal from the balanced photodetector was sent to the lock-in amplifier (MFLI 500 kHz, Zurich Instruments). Signals from both detectors were sent to the oscilloscope (MSO64B, Tektronix) and recorded for each data point. The sizes of the beams on the samples were measured using the Thorlabs beam profiler. The effective beam diameters were 400 µm and 70 µm for pump and probe, respectively. Both laser beams were focused on the sample which was placed inside a magneto-optical Cryostation (Montana Instruments). After the sample, the probe beam was guided onto the balanced photodetector.

Fig. S7 a) shows the transient Faraday angle as the function of the delay time for Mg-, Mn-, and pristine samples for the pump fluence of 900 µJ/cm². Fig. S7 b) shows the transient Faraday angle normalized by the material volume and absorption at the pump wavelength(515nm/2.41eV) (See Fig. S6) as the function of the delay time. We estimated the average thickness of the material from AFM data presented in supplementary note *"Microscopy characterization of the Mn-doped, Mg-doped and pristine MAPbBr₃ thin films",* while taking the size of the probe beam as the circle with a diameter of 70 µm. The pristine sample forms a nearly continuous polycrystalline thin film with an average thickness of about 200 nm. The Mn-doped sample shows an island-like morphology pattern with an average area coverage of about 20% and the average thickness of the islands is about 180 nm which is comparable to the pristine MAPbBr₃ sample. The Mg-doped sample shows 80 % coverage of the film with a thickness of about 300 nm. And for the sake of simplicity, we have taken the probing volume of the pristine sample equal to 1.

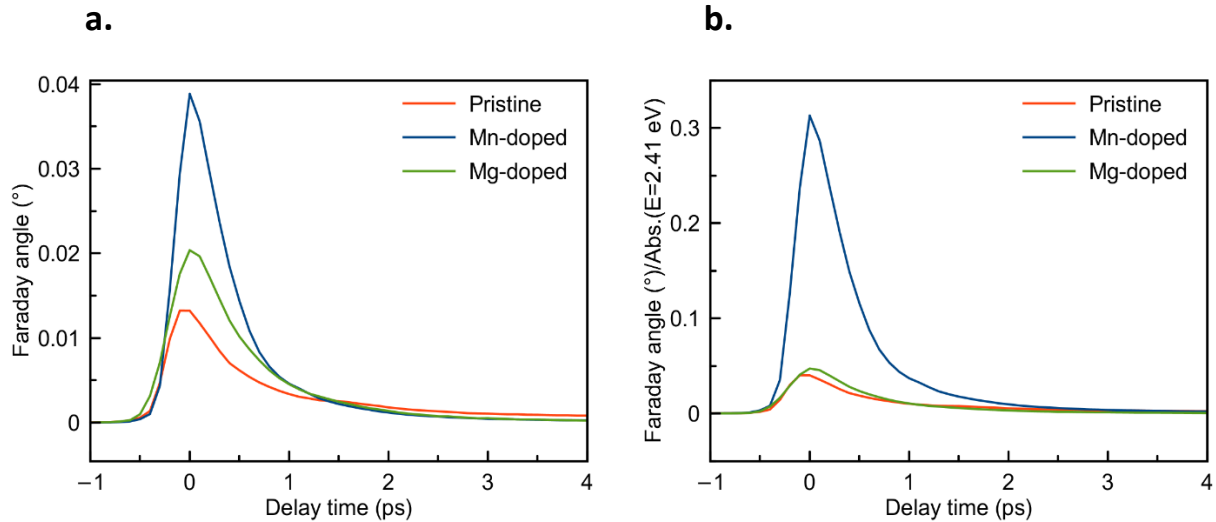

**Fig. S7. Transient Faraday signal for Mn-doped, Mg-doped and pristine samples**. *a) Transient Faraday angle measured at 4.6 K for Mn-doped, Mg-doped, and undoped MAPbBr₃ measured at a pump fluence of J=900 µJ/cm². b) Transient Faraday angle as a function of the delay time normalized by absorption at the energy corresponding to the pump energy of E=2.41 eV and the average volume of the material pro probe beam.*

### 6. Diagram of the optical transitions in the MAPbBr₃.

The valence band in lead halide perovskites consists of hybridized Pb s- and Br p-orbitals with the resulting s-like symmetry of the valence band while the conduction band consists of the hybridized Pb p-orbitals with Br s-orbitals with an overall resulting p-like symmetry. The presence of spin-orbit coupling and the Rashba effect leads to the splitting of the conduction band as shown in the diagram below. To fulfill the conservation of the total angular moment during the optical transition the condition $\Delta m_j = \pm 1$ must be fulfilled. Fig. S8 shows the schematic picture of the possible optical transitions induced by circularly polarized light in lead halide perovskites. Absorption of light with a certain helicity leads to the presence of an initial spin polarization level of 33% [1].

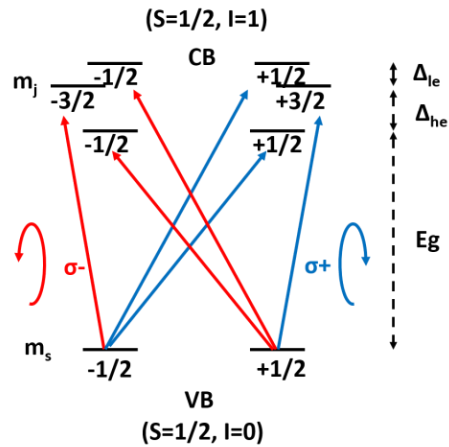

***Fig. S8. Optical transitions.*** *Schematic picture of the optical transitions in perovskites with an s-like valence and p-like conduction bands*

### 7. Transient absorption measurements.

To investigate the changes in charge carrier dynamics upon doping, we studied Mn-doped and pristine MAPbBr$_3$ thin films using transient absorption spectroscopy. TA experiments were performed using an optical setup in which samples were placed inside a Montana Cryostation, similar to the transient Faraday effect measurements, TA experiments were performed with the laser system (Pharos by Light Conversion) with a repetition rate of 5 kHz. The pump beam with 2.41 eV energy was guided through the delay stage while the broadband probe beam was guided directly onto the sample. After the sample, the probe beam was guided into a spectrometer with a mounted sCMOS camera (Kymera 193i and Zyla 5.5 by Andor Oxford Instruments). The sizes of the pump and probe beams were measured using a Thorlabs beam profiler.

Fig. S9 shows an example of the TA spectra for the pristine and Mn-doped MAPbBr$_3$ thin films measured at 5 K with linear excitations of the pump (2.41 eV) and probe pulses. The same energy for the pump pulse was used for the transient Faraday pump-probe experiment described in the main text. The value of the ground-state bleach was used in the main text for the calculation of the effective light-induced magnetization of Mn $M$(Mn) (Equation 1). The same curves were measured for a temperature range between 5 K and 200 K.

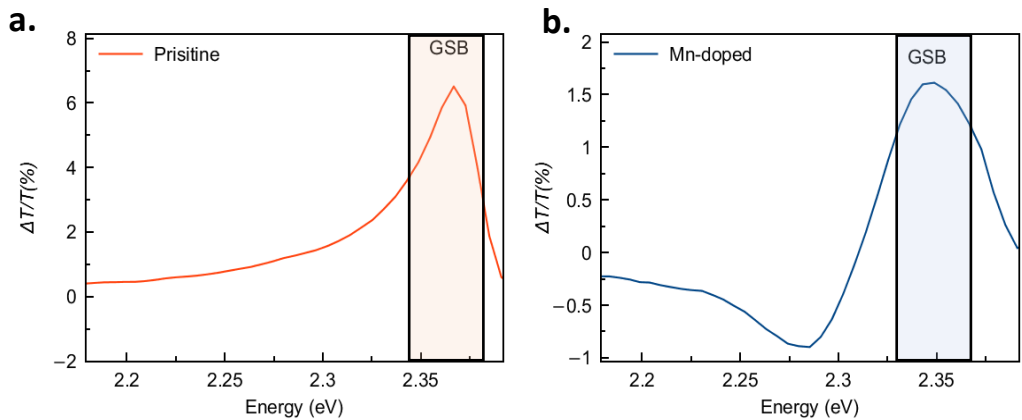

***Fig. S9. Transient absorption spectra.*** *a, b) Transient absorption spectra measured for linear pump and linear probe at the excitation density of 19 µJ/cm$^2$. measured for pristine and for 50% Mn-doped MAPbBr$_3$. The spectra were obtained for the delay time from 0 to 5 ps. Spectra were measured for the linear excitation of pump and probe beams. The filled area reflects the energy of the ground state bleach (GSB).*

## 8. Spin lifetime as the function of temperature und pump fluence.

To extract the spin lifetime, the data on the transient Faraday pump-probe signal were normalized by the maximum value, and the data for different temperatures and fluences were fitted with the single exponential decay law:

$$I(t) \sim e^{-t/\tau}$$

where $t$ is delay time, $\tau$ is spin life time.

Fig. S10 shows an example of the fitting of the experimental data on a transient Faraday pump-pump signal for 19.9 µJ/cm² measured at 15K with an exponential decay function.

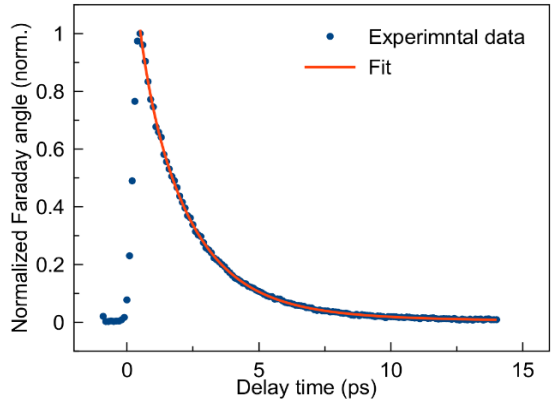

***Fig. S10. Transient Faraday angle fitting.*** *Transient Faraday angle as the function of delay time in Mn-doped thin film measured at 15K at pump fluence 19.9 µJ/cm². Blue dots reflect experimental data points. The red line shows fit using the exponential decay law.*

The same fitting routine was performed for the temperature region between 15K and 200K and for the pump fluence region between 5 µJ/cm² and 100 µJ/cm². Fig. S11. shows the extracted spin lifetime as the function of pump fluence and temperature.

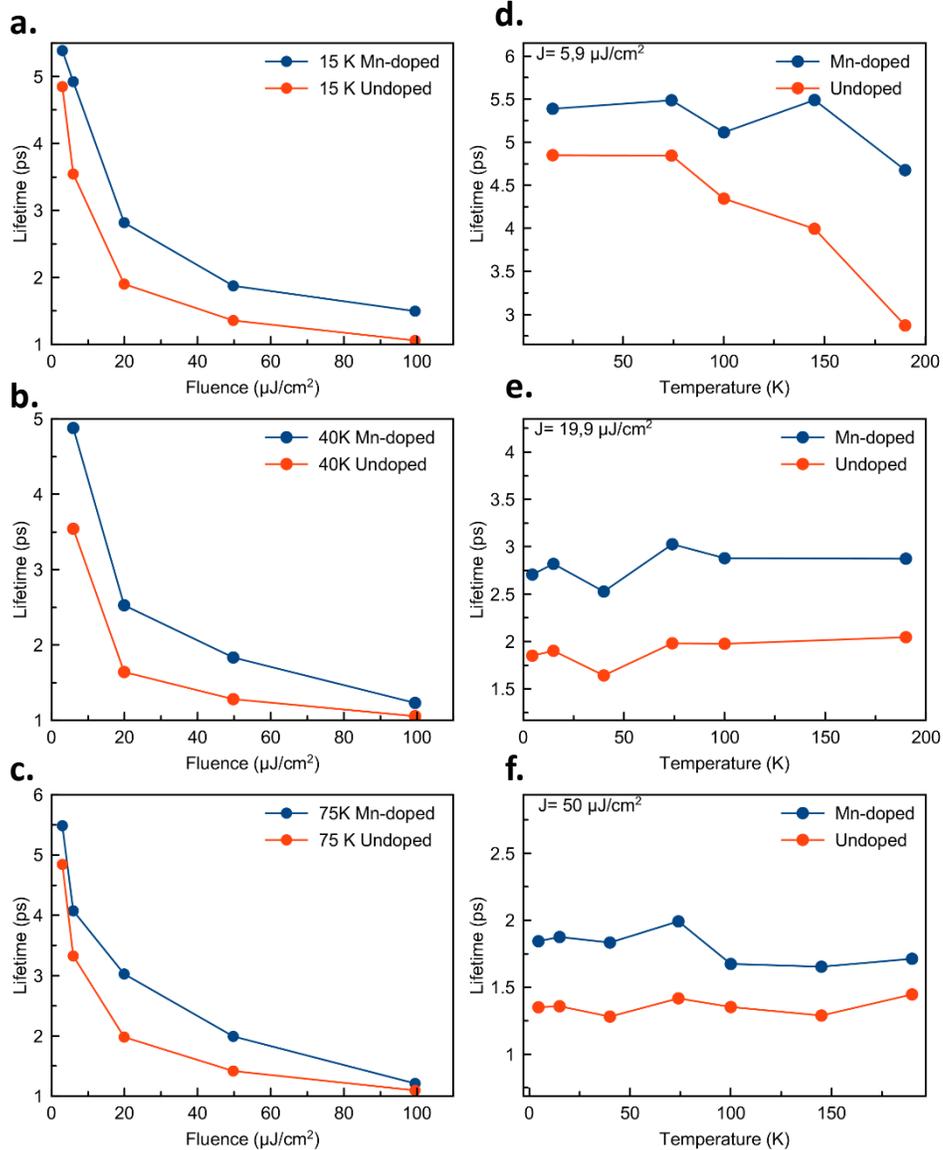

***Fig. S11. Spin life-time.*** *a, b, c) Spin lifetime of Mn-doped and pristine samples as the function of pump fluence at the temperatures 15K, 40K, and 75K respectively. d, e, f) Spin lifetime of Mn-doped and pristine samples as the function of temperature for different pump fluence of 6 µJ/cm$^2$, 20 µJ/cm$^2$, and 50 µJ/cm$^2$.*

## 9. Normalized Transient Faraday signal.

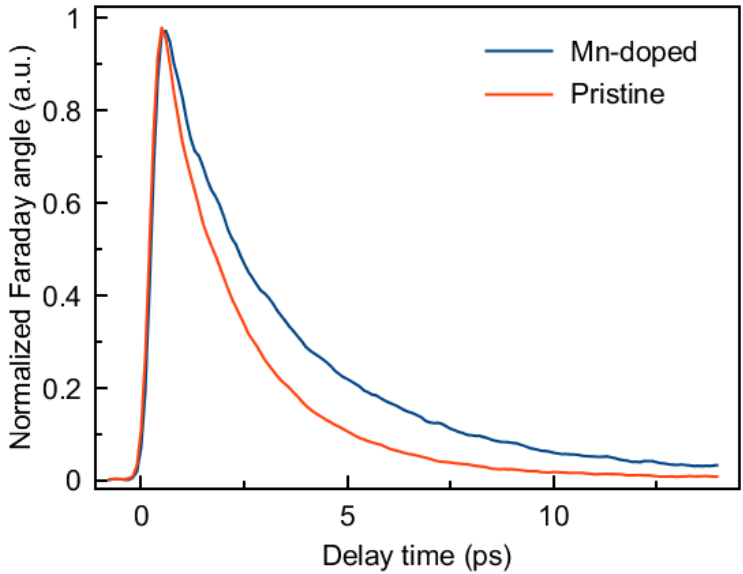

***Fig. S12. Normalized transient Faraday signal.*** *Normalized transient Faraday signal measured for Mn-doped and for the pristine MAPbBr$_3$ measured at the pump fluence 20 µJ/cm$^2$.*

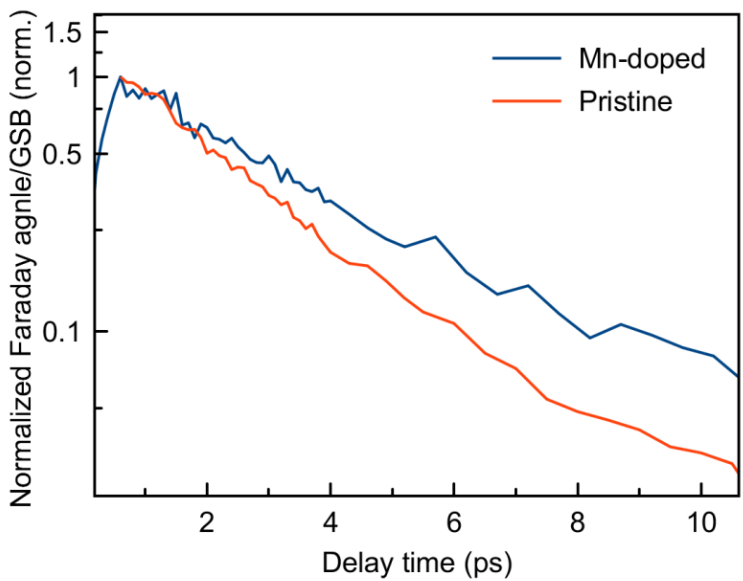

***Fig. S13. Transient Faraday signal normalized by the GSB dynamics.*** *Transient Faraday signal for Mn-doped and for the pristine MAPbBr$_3$ measured for the pump fluence 20 µJ/cm$^2$ normalized by the ground state bleach dynamics measured under linear excitation at the same pump fluence.*